\documentclass[notitlepage, showpacs, nobalancelastpage, twocolumn, aps, prl, 10pt, floatfix]{revtex4-2}
\usepackage[english]{babel}
\RequirePackage[T1]{fontenc}
\RequirePackage{times}

\usepackage{siunitx}[=v2] %
\usepackage[italicdiff]{physics}
\usepackage{amsfonts}
\usepackage{subfigure}
\usepackage{amsmath}
\usepackage{amssymb} 
\usepackage{graphicx,epstopdf}
\usepackage{appendix}
\usepackage{xspace}
\usepackage{lipsum}
\usepackage{array}
\usepackage{hyperref}
\usepackage[dvipsnames]{xcolor}
\usepackage{my_symbols}
\usepackage{CJK}
\usepackage{bm}
\usepackage{verbatim}
\usepackage{tikz}
\usepackage{booktabs}
\usepackage{multirow}
\usepackage{pifont}
\usepackage[normalem]{ulem}
\usepackage{threeparttable}
\usepackage{setspace}

\begin{document}

\begin{CJK*}{UTF8}{gbsn}
	\title{Geometric phase and a nonreciprocal spin wave circular polarizer}
	\author{Yu Liu (刘瑜)}
	\author{Jin Lan (兰金)}
	\email[Corresponding author:~]{lanjin@tju.edu.cn}
	\affiliation{Center for Joint Quantum Studies and Department of Physics, School of Science, Tianjin University, 92 Weijin Road, Tianjin 300072, China}
	\affiliation{Tianjin Key Laboratory of Low Dimensional Materials Physics and Preparing Technology, Tianjin University, Tianjin 300354, China}
	\begin{abstract}
		We show that spin wave acquires a polarization-dependent geometric phase along a cyclic trajectory of non-coplanar magnetizations in antiferromagnets. 
		Specifically, we demonstrate that a cyclic set of $90^\circ$ antiferromagnetic domain walls simultaneously introduce geometric and dynamic phases to spin wave, and thus leads to asymmetric magnitude of overall phase for left-/right-circular components. 
		Based on the polarization-dependent  phase, we propose theoretically and confirm by micromagnetic simulations that, a Mach-Zehner interferometer with cyclic $90^\circ$ domain walls in one arm and homogenous domain in the other arm,  naturally acts as a spin wave circular polarizer. 
		Moreover, the circular polarizer has intrinsic nonreciprocity, which filters opposite polarization in opposite propagation direction.
	\end{abstract}
	\maketitle
\end{CJK*}

\emph{Introduction.}
Phase is the core property of all waves, including electromagnetic wave, acoustic wave, matter wave, gravitational wave as well as spin wave. 
The wave phase naturally divides into two parts: the dynamical phase characterizing the wave evolution rate, and the geometric phase describing the geometric property of the wave system in parametric space \cite{berry_quantal_1984,bhandari_polarization_1997,xiao_berry_2010}.
Since its initial proposal, the concept of geometric phase has fastly evolved and become the foundation of vast and diverse disciplines \cite{berry_geometric_2010,cohen_geometric_2019}.
Exploitation of geometric phase offers new possibilites in wave manipulation including trajectory controlling, wavefront tailoring and polarization harnessing, beyond the physical limit imposed by the dynamical phase \cite{bliokh_spin_2015,jisha_geometric_2021,roux_geometric_2006,kim_fabrication_2015,arbabi_dielectric_2015,xiao_geometric_2015,slussarenko_guiding_2016,zhu_wavevectorvarying_2021}.

Spin wave, the collective precession of ordered magnetization, is an alternative angular momentum carrier beside the spin-polarized conduction electron \cite{kajiwara_transmission_2010,chumak_magnon_2015,cornelissen_longdistance_2015,barman_2021_2021}. 
The dynamical phase of spin wave can be tuned via multiple means, such as exerting magnetic field \cite{kostylev_spinwave_2005,schneider_realization_2008} or electric field \cite{liu_electric_2011,zhang_electricfield_2014,cheng_antiferromagnetic_2016}, passing electric current, placing magnetic impurities \cite{dobrovolskiy_spinwave_2019,yu_magnetic_2020,wang_inversedesign_2021}, coupling between two waveguides \cite{wang_reconfigurable_2018,wang_magnonic_2020,zhao_reconfigurable_2022}, and deposting magnetic domain walls \cite{hertel_domainwall_2004,lan_antiferromagnetic_2017,han_mutual_2019,ye_magnetically_2021}.
Based on the dynamical phase shift, a plethora of logic and neuromorphic magnonic devices have been theoretically proposed or experimentally realized \cite{schneider_realization_2008,papp_nanoscale_2021,pirro_advances_2021,chumak_advances_2022}.

In contrast to extensively investigated dynamical phase, the geometric phase of spin wave is only studied in limited situations. 
The geometric phase is shown to develop in a magnetic ring \cite{dugaev_berry_2005}, between two magnetic domain walls \cite{buijnsters_chiralitydependent_2016} or along a magnetic helix \cite{wu_curvilinear_2022} where non-coplanar magnetization forms along the spin wave trajectory. 
However, systematic formulation of the geometric phase for spin wave is still lacking, impeding the full exploitation of geometric phase  in design of magnonic devices, let alone the collaborative leverage of geometric and dynamical phases.

In this work, we show that spin wave acquires a geometric phase cross a cyclic set of non-coplanar $90^\circ$ domain walls, beside the conventional dynamical phase.
By virtue of the polarization-dependent geometric phase, we propose a spin wave circular polarizer, based on the wave interference in a two-arm Mach-Zehner structure.
We further show that functionality of the circular polarizer is highly reprogrammable by reversing the propagation direction, tuning the working frequency, or altering the magnetic states.
Parallel wave processings boosted by the fundamental superposition principle, are also demonstrated upon such a circular polarizer.

\begin{figure}[b]
	\centering
	 {\includegraphics[width=0.48\textwidth, trim=0 0 10 10,clip]{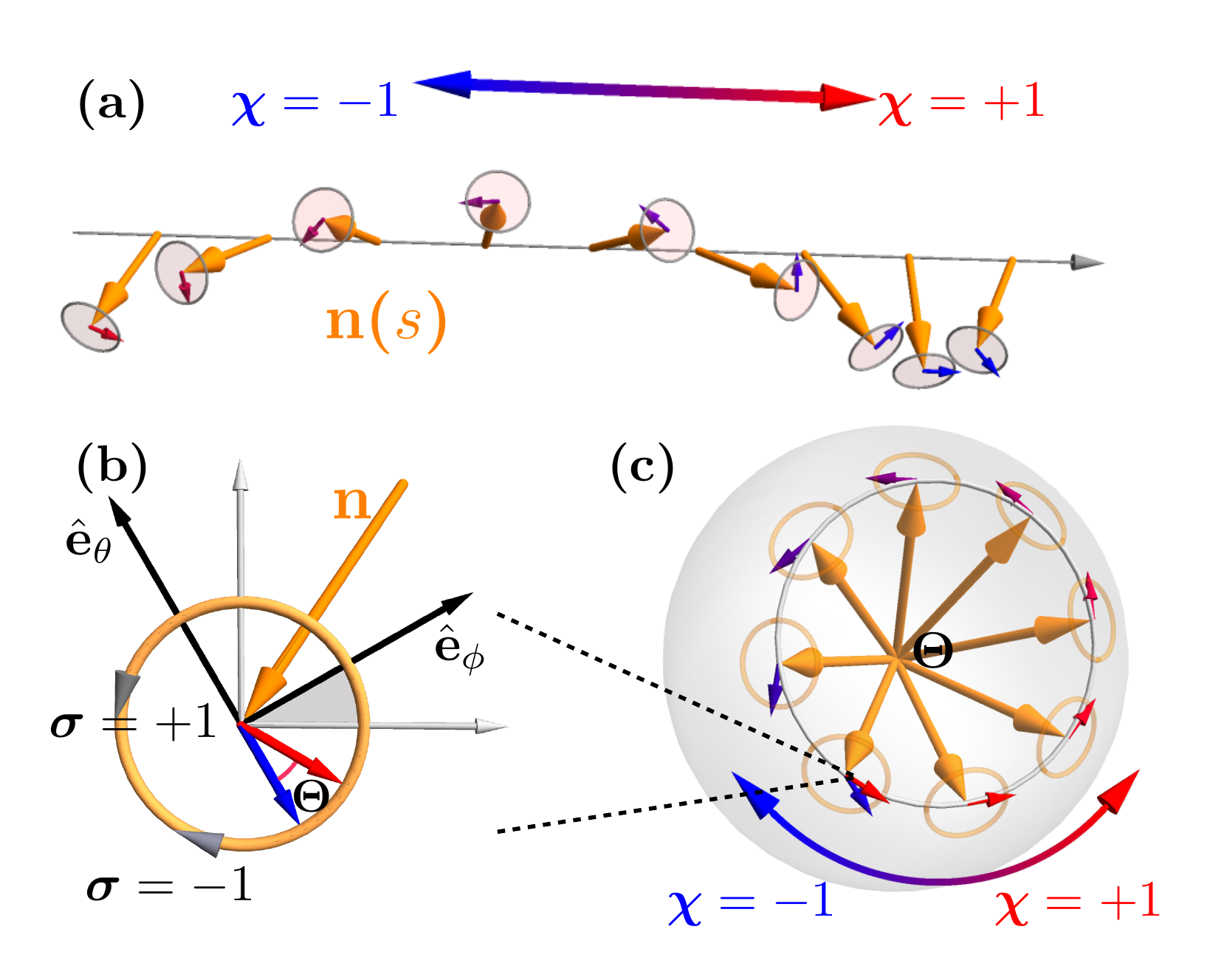}} 
	
	\caption{ {\bf Schematics of the geometric phase acquired by polarized spin wave in non-coplanar magnetizations.}
		(a) The evolution of the polarization direction along a closed trajectory of non-coplanar magnetizations. 
		(b) The geometric phase of circular polarization and the rotation of linear polarization.  
		(c) The evoluation of the polarization direction in a magnetic Bloch sphere. 
		The orange arrows are for background N\`eel order $\bn$, and the accompanying red/blue arrows are for a specific polarization direction along trajectory.
		\label{fig:geometric}}
\end{figure}

\emph{Polarized spin wave in antiferromagnets.}
Consider an antiferromagnet with the normalized N\`eel order denoted by unit vector $\bN$, which is naturally partitioned to the static background magnetization $\bn$ and the dynamical spin wave $\bn'$, $\bN=\bn+\bn'$.
Under the unity constraints $|\bN|=1$ and $|\bn|=1$, as well as the small amplitude approximation for spin wave $|\bn'|\ll 1$, the transverse condition $\bn\cdot \bn'=0$ is satisfied everywhere.
Hence, it is instructive to formulate spin wave in spherical coordinates as $\bn'=n_\theta \hbe_\theta +n_\phi \hbe_\phi$, where  $\hbe_{\theta/\phi}$ are two orthgonal polarization directions transverse to the background  magnetization $\hbe_r\equiv\bn$, and $n_{\theta/\phi}$ are the corresponding wave components.
Alternatively, in complex form, the spin wave reads $\bn'=\sum_{\sigma=\pm 1} \psi_\sigma \bm{\xi}_\sigma $, where $\bm{\xi}_\sigma=(\hbe_\theta+i\sigma\hbe_\phi)/2$ and $\psi_\sigma=n_\theta-i \sigma n_\phi$ with $\sigma=\pm 1$ are bases and components of the left-/right-circular polarizations, respectively.

\emph{Geometric phase of polarized spin wave.}
The $SO(2)$ symmetry of the linear bases $\hbe_{\theta/\phi}$ about the background magnetization, gives rises to the $U(1)$ symmetry of the circular bases $\bm{\xi}_{\pm}$, or an indeterminate phase for the circular polarization.
Hence, when a polarized spin wave travels along a closed trajectory of inhomogeneous magnetization $\bn(s)$ parametrized by the arc length $s$, the circular basis may develop an additional geometric phase instead of restoring to its original phase.
Specifically, the relative phase of spin wave developed between $\bn$ and $\bn+d\bn$ is characterized by the Berry connection $\bm{\Lambda}_{\sigma\sigma'}=-i \bm{\xi}_{\sigma}^\dag \cdot \nabla_\bn \bm{\xi}_{\sigma'}$, which is diagonal in circular bases with $\bm{\Lambda}_{\sigma\sigma'}=\delta_{\sigma\sigma'} \bm{\Lambda}_\sigma$ \cite{onoda_hall_2004}. 
The accompanying Berry curvature is $\bm{\Omega}_\sigma=\nabla_\bn \times \bm{\Lambda}_\sigma= \sigma \bn$, resembling the field radiated from a monopole of strength $\sigma$ located at $\bn=0$.
The evolution of the circular bases is then governed by $\partial_s \bm{\xi}_\sigma=-i (\bm{\Lambda}_\sigma \cdot \partial_s \bn )\bm{\xi}_\sigma  $ with solution $\bm{\xi}_\sigma= \exp(i\Phi^G_{\sigma\chi}) \bm{\xi}_\sigma^0$, where $\bm{\xi}_\sigma^0$ is the initial circular basis, and $\chi$ denotes the propagation direction along the trajectory in Fig. \ref{fig:geometric}(a).
The geometric phase accumulated in a closed trajectory of background magnetization is thus described by
\begin{align}
	\label{eqn:phase_geo}
	\Delta \Phi^G_{\sigma\chi}= -\oint_l \bm{\Lambda}_\sigma \cdot d\bn
	 = -\sigma \chi \Theta,
\end{align}
where $\Theta$ is the magnitude of solid angle enclosed by trajectory $l$
in a magnetic Bloch sphere, and $\chi$ corresponds to anticlockwise/clockwise circulating direction, as depicted in Fig. \ref{fig:geometric}(c).
The geometric phase in \Eq{eqn:phase_geo} shares a similar form to the spin-redirectional phase in its optical counterpart \cite{onoda_hall_2004, bliokh_spin_2015, bliokh_geometric_2019}, but the solid angle is subtended by background magnetizations here instead of optical wavevectors.

In \Eq{eqn:phase_geo}, the geometric phase $\Delta \Phi^G_{\sigma\chi}$ flips sign when either the trajectory reverses its direction ($\chi\to -\chi$) or the spin wave alters its circular polarization ($\sigma\to-\sigma$), indicating its intrinsic chirality. 
Moreover, opposite geometric phase $\pm \Theta$ experienced by two circular modes leads to Faraday rotation of the linear bases, 
\begin{align}
	\label{eqn:pol_rot}
	\begin{pmatrix} 
		\hbe_{\theta} \\
		\hbe_{\phi}
	\end{pmatrix} =  
	\begin{pmatrix} 
		\cos\Theta  & \chi \sin\Theta  \\
		-\chi \sin\Theta & \cos\Theta
	\end{pmatrix}
	\begin{pmatrix} 
		\hbe_{\theta}^0 \\
		\hbe_{\phi}^0
	\end{pmatrix},
\end{align}
where $\hbe_{\theta/\phi}^0$ are the initial direction.

\emph{Geometric phase across $90^\circ$ domain walls.}
To elaborate the concept of geometric phase, we turn to an antiferromagnet wire   with magnetic cubic anisotropy \cite{ye_magnetically_2021}.
The dynamics of the N\`eel order $\bN$ is governed by antiferromagnet-type Landau-Lifshitz-Gibert (LLG) equation \cite{haldane_nonlinear_1983,tveten_antiferromagnetic_2014,wu_curvilinear_2022},
\begin{align}
	\label{eqn:LLG}
	\rho \bN\times\ddot{\bN}= -\bN \times \gamma \bH+ \alpha \bN \times \dot{\bN},
\end{align}
where $\rho$ is the inertia of antiferromagnetic dynamics, $\gamma$ is the gyromagnetic ratio, and  $\alpha$ is Gilbert damping constant.
Here $\bH= -\delta U/\delta \bN$ is the effective field acting on N\`eel order $\bN$, $U=(1/2)\int \qty[A (\nabla\bN)^2+ K (N_x^2 N_y^2+N_y^2N_z^2+N_x^2N_z^2)]dx$ is the magnetic energy, $A$ is the exchange stiffness, $K$ is the cubic anisotropy strength.
The inertia is expressed by $\rho=a^2/8\gamma A$, where $a$ is the lattice constant.
The dipolar field and a moderate easy-axis anisotropy do not change the main physics in this work, and thus are disregarded \cite{SM}.

\begin{figure*}[bt]
	\centering
	{\includegraphics[width=0.99\textwidth, trim=0 0 0 20,clip]{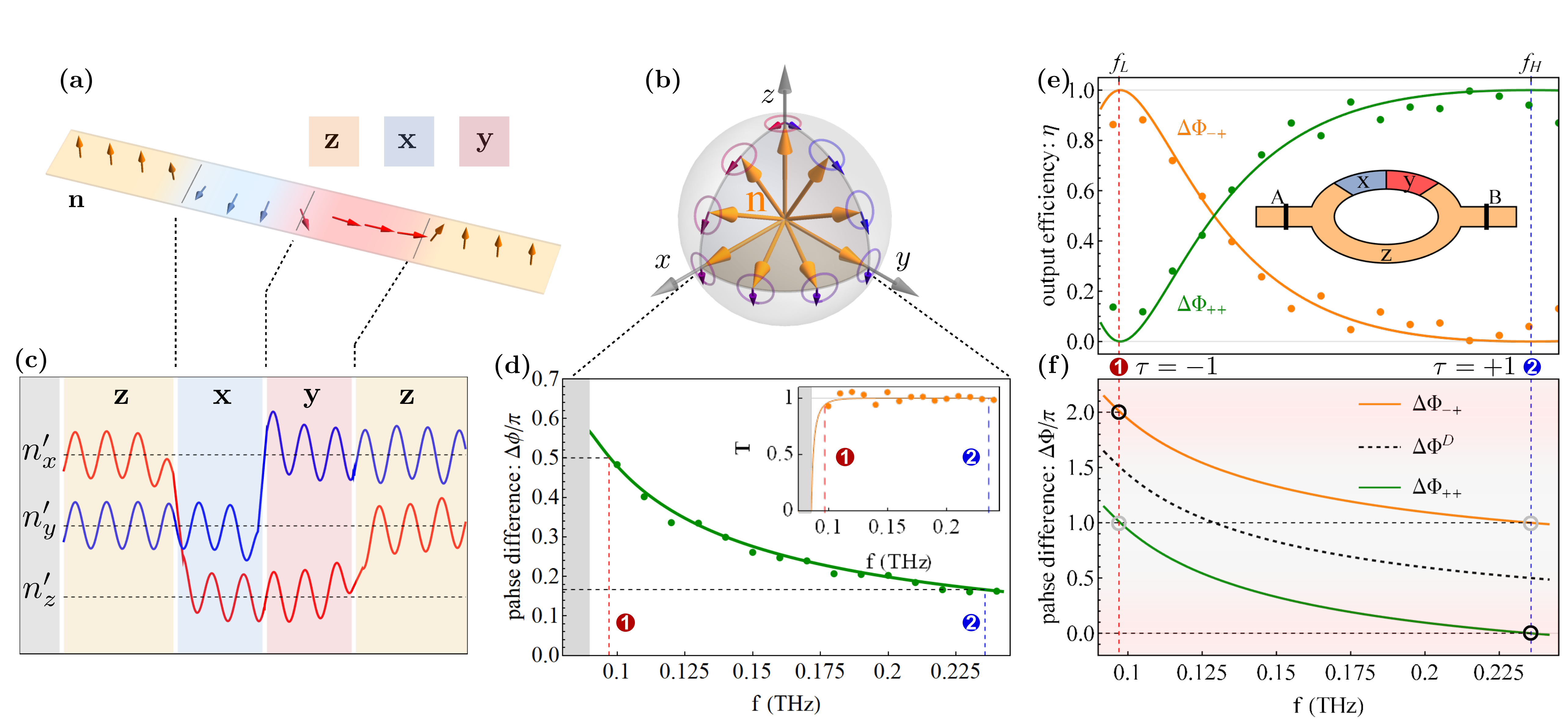}}
	\caption{
		{\bf Spin wave scattering and interference across the cyclic $90^\circ$ domain walls.}
		(a) Magnetic profile of the cyclic $\hbz$-$\hbx$-$\hby$-$\hbz$ domain walls. 
		(b) Parallel transportation of a polarization direction in a magnetic Bloch sphere. 
		(c) Braiding of linear-$x/y$ polarizations across three $90^\circ$ domain walls extracted from micromagnetic simulations.
		(d) Dynamical phase across a single $90^\circ$ domain wall. 
		The inset plots the transmission probability of spin wave. 
		(e) The output efficiency of two circular polarizations. 
		The inset depicts the schematics of the two-arm interferometer and the corresponding magnetic profiles.
		(f) The overall phase for left-/right circular polarizations.
        In (d)(e)(f), the solid lines are theoretical calculations based on \Eq{eqn:sw_eom}, and the dots are extracted from micromagnetic simulations, and  two circled marks are for low/high frequency.		\label{fig:phase}
	}
\end{figure*}

Due to the cubic anisotropy, the magnetization direction of a homogenous domain lies at one of the three Cartesian directions $\bn=\hbx_i$ with $i=\qty{1,2,3}$.
When two orthogonally magnetized $\hbx_i$-domain  and $\hbx_j$-domain meet, a $90^\circ$ domain wall forms with all magnetizations residing in the $x_i$-$x_j$ plane.
The non-coplanar magnetizations in three cyclically connected $\hbx_i$-$\hbx_j$-$\hbx_k$-$\hbx_i$ domain walls subtends a solid angle of exactly $\Theta=\pi/2$ in magnitude, as depicted in Fig. \ref{fig:phase}(b).
Therefore, the linear-$x/y$ modes interchanges after traversing  such a cyclic domain wall according to \Eq{eqn:pol_rot},  or following the parallel transport law in a magnetic Bloch sphere as depicted in Fig. \ref{fig:phase}(b).

The spin wave evolution are further investigated by micromagnetic simulations in Mumax3 \cite{vansteenkiste_design_2014} with  following magnetic parameters: the exchange coupling constant $A=\SI{2.1e-11}{J/m}$, the gyromagnetic ratio $\gamma=\SI{2.21e5}{m/(A.s)}$, the cubic anisotropy $K=\SI{1.0e4}{J/m^2}$, the damping constant $\alpha=\SI{1.0e-5}{}$, and the lattice constant $a=\SI{0.5}{nm}$.
An $\hbx$-domain and a $\hby$-domain are placed in between $\hbz$-domains at two sides of an antiferromagnetic wire in Fig. \ref{fig:phase}(a), and linearly polarized spin waves are injected from the left side. 
As shown in Fig. \ref{fig:phase}(c), after traversing two intermediate $\hbx$- and $\hby$- domains, linear-$x/y$ spin waves are braided by exploiting linear-$z$ as the third state, similar to  its optical and acoustic counterparts \cite{iadecola_nonabelian_2016,zhang_nonabelian_2022,chen_classical_2022}.
Nevertheless, non-Abelien braiding is absent since only two independent polarization modes exist upon any background magnetization.
We have checked that the polarization braiding, as a special case of polarization rotation, is independent of spin wave frequency.
In contrast, the linear-$x/y$ spin wave remains unchanged for a trivial trajectory through the $\hbz$-$\hbx$($\hby$)-$\hbz$ domains, since the enclosing solid angle is zero therein \cite{SM}.

\emph{Dynamical phase across a $90^\circ$ domain wall.}
Beside introducing geometric phase, the domain wall also modifies the dynamical phase by altering the spin wave dynamics \cite{hertel_domainwall_2004,buijnsters_chiralitydependent_2016,han_mutual_2019}. 
Without loss of generality, we consider a $90^\circ$ domain wall lying between a $\hbx$-domain and a $\hby$-domain, which adopts a Walker-type profile $\bn(x)=(\sqrt{[1-\tanh(x/W)]/2},\sqrt{[1+\tanh(x/W)]/2},0)$ with $W=\sqrt{A/K}$ the characteristic width \cite{tveten_Staggered_2013,ye_magnetically_2021}. 
The spin wave dynamics upon such an $\hbx$-$\hby$ type  $90^\circ$ domain wall is then recast from the LLG equation \eqref{eqn:LLG} to a Klein-Gordon-like equation 
\begin{align}
	\label{eqn:sw_eom}
	-\frac{\rho}{\gamma} \ddot{\psi} = \qty[-A\partial_x^2 +K+ V(x) ]\psi,
\end{align}
where $V(x)=-(11/8)K\sech^2(x/W)$ is the effective potential well caused by the inhomogeneous magnetization within domain wall.
The deviation of $V(x)$ from the celebrated P\"oschl-Teller type potential \cite{lan_antiferromagnetic_2017,yu_polarizationselective_2018} originates from the additional contribution of the cubic anisotropy.
The hybridization of two circular modes caused by  the cubic anisotropy, or the retarding effect between two linear modes \cite{ye_magnetically_2021,SM}, is disregarded in \Eq{eqn:sw_eom}  for model simplicity and compactness.

The spin wave scatterings by a $90^\circ$ domain wall are quantitatively investigated via  two numerical evaluations in parallel: the Green function calculations based on \Eq{eqn:sw_eom} via Kwant \cite{groth_kwant_2014} and micromagnetic simulations via Mumax3 \cite{vansteenkiste_design_2014}. 
The agreements between two methods in Fig. \ref{fig:phase}(d) corroborate the following two influences caused by the potentail well $V(x)<0$ or the inhomogeneous domain wall profile $\bn(x)$: 
i) A small reflection in the extremely low frequency range; ii) A positive dynamical phase $\Delta\phi>0$ in the full range, which monotonically decreases for increasing frequency.

\begin{figure*}[bt]
	\centering
	{\includegraphics[width=0.99\textwidth, trim=0 0 0 0,clip]{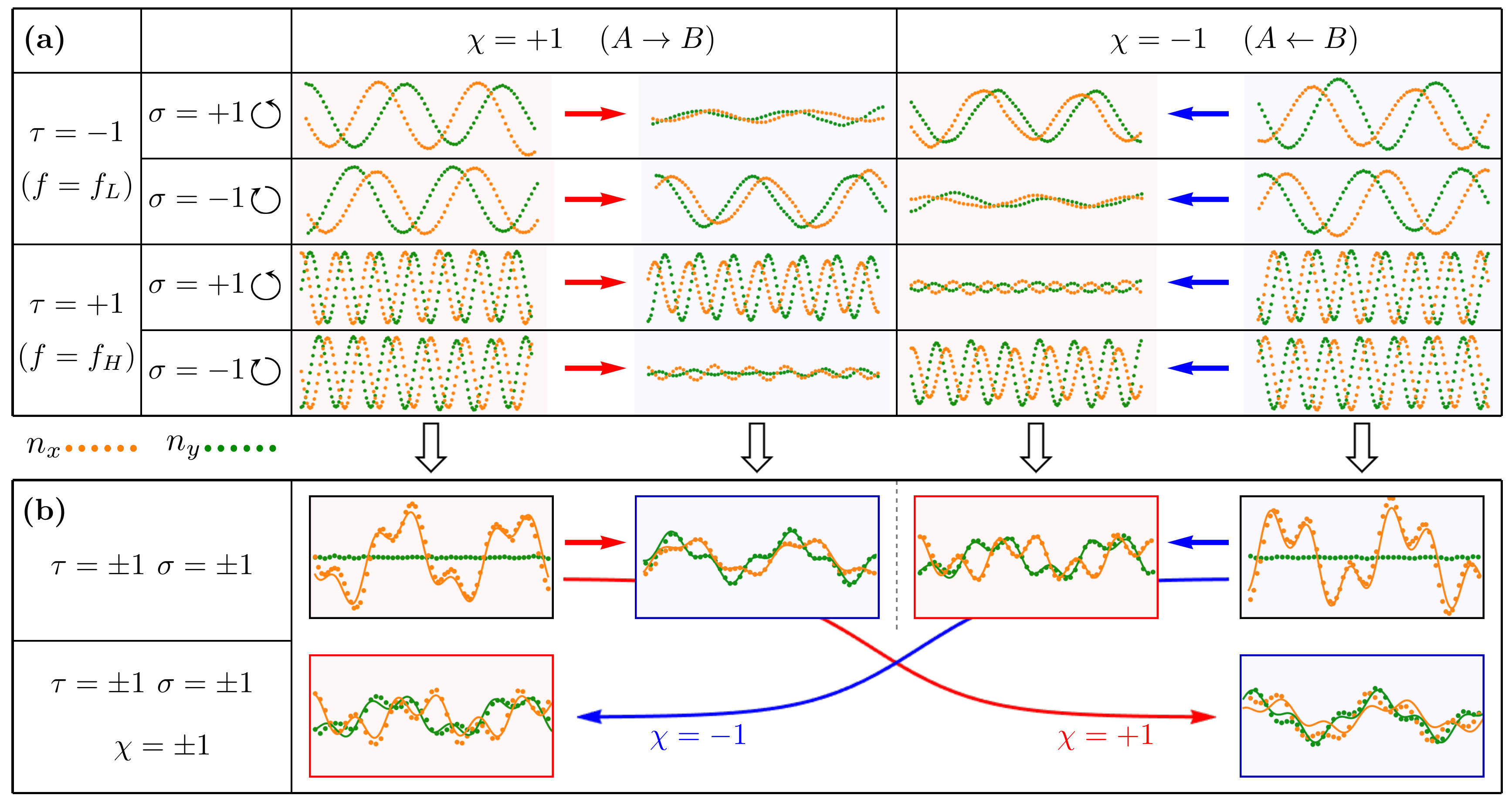}}
	\caption{
		{\bf Spin wave processing in a non-reciprocal circular polarizer.}
		(a) Individual processing in a single channel. 
		(b) Parallel processing in multiple channels. 
		Upper pannel: unidiretional processing in $2^2$ channels of $\qty{\sigma,\tau}$ separately for $\chi=1$ and $\chi=-1$;
		Lower panel: bidirectional processing of $2^3$ channels of $\qty{\sigma,\tau,\chi}$ simultaneously for $\chi=\pm 1$. 
		In all plots, the orange/green dots are for $n_{x/y}$ extracted from micromagnetic simulations, and the solid lines are for theoretical fittings.
	}
	\label{fig:circular_polarizer}
\end{figure*}

\emph{Overall phase across cyclic $90^\circ$ domain walls.}
Consider a Mach-Zehner type spin wave interferometer as depicted in Fig. \ref{fig:phase}(e) inset, where a $\hbz$-domain occupies the major region, and a $\hbx$-domain and a $\hby$-domain are deposited in the upper arm.
When  spin waves split from the input port converge again at the output port, the overall phase difference developed between the cyclic $\hbz$-$\hbx$-$\hby$-$\hbz$ domain walls in the upper arm and the homogenous $\hbz$-domain in the lower arm is  described by
\begin{align}
	\label{eqn:total_phase}
	\Delta\Phi_{\sigma\chi}=\Delta\Phi_{\sigma\chi}^G+\Delta\Phi^D=-\frac{\sigma\chi\pi}{2}+3\Delta\phi,
\end{align}
where the geometric part $\Delta\Phi_{\sigma\chi}^G$ arises from the enclosing solid angle $\Theta=\pi/2$, and the dynamical part $\Delta\Phi^D$ is accumulated across three consecutive domain walls.

The frequency-dependent overall phase $\Delta\Phi^D$  in Fig. \ref{fig:phase}(f) guide us to designate binary parameter $\tau$ to following two specific frequencies: $\tau=-1$ for low frequency $f_L\approx\SI{0.097}{THz}$ with $\Delta\Phi^D=3\pi/2$;
and $\tau=+1$ for high frequency $f_H\approx\SI{0.236}{THz}$ with $\Delta\Phi^D=\pi/2$.
The manipulation space $\qty{\sigma,\chi,\tau}$ formed by polarization states in $\sigma$, propagation direction in $\chi$ and freqeuncy in $\tau$ then provides $3$ binary means to harness spin wave.
In such a $3$-bit manipulation space, the overall phase is recast from \Eq{eqn:total_phase} to 
\begin{align}
	\label{eqn:total_phase_new}
	\Delta\Phi_{\sigma\chi\tau}=\Delta\Phi_{\sigma\chi}^G+\Delta\Phi^D_\tau=\frac{2-\tau-\sigma\chi}{2}\pi,
\end{align}
which is always integer mutiples of $\pi$.
Consequently,  the interference of spin waves in two arms at the confluence region leads to the output efficiency
\begin{align}
	\label{eqn:sw_int}
      \eta_{\sigma\chi\tau}  %
	    = \cos^2\frac{\Delta\Phi_{\sigma\chi\tau}}{2}= \frac{1+\sigma\chi\tau}{2},
\end{align}
which lies in a binary on/off states, i.e., being either completely constructive $\eta=1$ for $\Delta\Phi=0, 2\pi$, or completely destructive $\eta=0$ for $\Delta\Phi=\pi$, as depicted in Fig. \ref{fig:phase}(e).
It is noteworthy that the on/off states possess a rather wide frequency tolerance, especially for the high-frequency case.

\emph{A non-reciprocal circular polarizer.} 
In \Eq{eqn:sw_int}, the on/off combinations arises in the interferometer for arbitrary binary pair in $\sigma$, $\chi$ or $\tau$. 
One circular polarization is blocked $\eta_\sigma=0$ while the other is passed $\eta_{-\sigma}=1$, hence the two-arm interferometer acts as a spin wave circular polarizer. 
And when the propagation direction is reversed ($\chi\to-\chi$), the on/off state is flipped, indicating the non-reciprocity of the circular polarizer. 
The chirality of such a circular polarizer originates from the chiral nature of the geometric phase as manifested in \Eq{eqn:phase_geo}.
By switching between high/low frequency ($\tau\to-\tau$), the on/off state is flipped again, revealing a frequency-controlled chirality of the circular polarizer.

The polarization-filtering functionality of the circular polarizer is further confirmed by micromagnetic simulations in Fig. \ref{fig:circular_polarizer}(a). 
Circular spin waves at selected frequencies are excited at $A$ or $B$ port of the interferometer in Fig. \ref{fig:phase}(e), and is detected in the other port.  
At low working frequency $f=f_L$, when two circular spin waves are continuously excited at port A, only the left-circular mode is observed at port $B$, signifying a left-circular polarizer in the rightward direction. 
In contrast, only the left-circular mode excited at port $B$ is visible at port $A$, suggesting a right-circular polarizer in the leftward propagation direction. 
After switching to the high working frequency $f=f_H$, the right/left-circular mode is selectively passed for $A\to B$ ($A\leftarrow B$), in full compliance to the on/off rule outlined in \Eq{eqn:sw_int}.

\emph{Superposition-endorsed parallel wave processing.}
The principle of wave superposition not only engenders the interference for waves of the same type, but also ensures that waves of different types can be brought to the same medium without mutual disturbances in linear regime \cite{goldstein_polarized_2017}.
Guided by this fundamental superposition principle, individual manipulations in mono-polarized, unidirectional and monochromatic fashion in Fig. \ref{fig:circular_polarizer}(a) can be processed concurrently in the interferometer at once.
In Fig. \ref{fig:circular_polarizer}(b), when a linear-polarized and dichromatic spin wave consisting of equal components of high/low frequencies and left-/right-circular polarizations is continuously injected from port $A$, only the low-frequency left-circular and the high-frequency right-circular components is detected at port $B$.
Similarly, when port $B$ is set as the input port, the other two components are detected in port $A$.
In both cases ($\chi=\pm 1)$, the input and output signals in Fig. \ref{fig:circular_polarizer}(b) are simple addition of $2^2$ channels in $2$-bit $\qty{\sigma,\tau}$ space in Fig. \ref{fig:circular_polarizer}(a).

To fully unleash the power of parallel processing, the linear-polarized and dichromatic spin wave are simultaneously excited at port $A$ and $B$ for finite duration of $\SI{0.16}{ns}$. 
These two wave pulses travelling in opposite directions then encounter and penetrate each other, and are simultaneously detected at the other side after a waiting time $\SI{0.20}{ns}$.
Aided by the wave pulse, a parallel processing of $2^3$ channels for $3$-bit $\qty{\sigma,\chi,\tau}$ space is enabled without much signal deterioration  in this bidirectional setup, as  also demonstrated in Fig. \ref{fig:circular_polarizer}(b).

\emph{Magnetic programmability and scalability.}
By altering the magnetic states in the interferometer, the geometric/dyanmical phase can be modified, and the functionality of the circular polarizer adjusts accordingly \cite{SM}.
When any magnetic domain is switched $\hbx_i\to -\hbx_i$,  the geometric phase $\Delta\Phi^G$ alters its sign, and thus the circular polarizer changes its chirality.
Meanwhile, when magnetic domain walls along the ring formed by two arms is moved, the dynamical phase becomes $\Delta\Phi^D= d \Delta\phi$ with $d=\qty{-3,-1,1,3}$ the number difference of domain wall in two arms, the circular polarizer at low frequency $f=f_L$ changes its filtering chirality. 

The circular polarizer in this work is fully captured by two scales: the typical length $W=\sqrt{A/K}$ and the typical time $t_0=(\pi\gamma/\sqrt{2}a)\sqrt{AK} $. %
Hence, the whole design  can be directly scaled for any combination of the exchange stiffness $A$ and cubic anisotropy $K$.

\emph{Conclusion.}
In conclusion, we show that spin wave acquires a geometric phase along a cyclic trajectory of non-coplanar magnetizations in antiferromagnets. 
Moreover, a cyclic set of $90^\circ$ domain walls imposes a polarization-dependent phase to passing spin wave, which consists of both geometric and dynamic phase. 
By virtue of such asymmetric phase for left/right-circular modes, we realize an interference-based circular polarizer.
The collaboration between geometric and dynamical phase, provides new paradigms in harnessing spin wave via non-coplanar magnetizations.

\emph{Acknowledgements.}
J.L. is grateful to Jiang Xiao for insightful discussions.
This work is supported by National Natural Science Foundation of China (Grant No. 11904260) and Natural Science Foundation of Tianjin (Grant No. 20JCQNJC02020).


%

\end{document}